# Comment on "Neutron diffraction evidence of the 3-dimensional structure of Ba$_2$MnTeO$_6$ and misidentification of the triangular layers within the face-centred cubic lattice"

J. Khatua[1], T. Arh[2,3], Shashi B. Mishra[4], H. Luetkens[5], A. Zorko[2,3,†], B. Sana[1], M. S. Ramachandra Rao[6], B. R. K. Nanda[4,‡], and P. Khuntia[1,7,*]

[1]*Department of Physics, Indian Institute of Technology Madras, Chennai 600036, India*
[2]*Jožef Stefan Institute, Jamova c. 39, 1000 Ljubljana, Slovenia*
[3]*Faculty of Mathematics and Physics, University of Ljubljana, Jadranska u. 19, 1000 Ljubljana, Slovenia*
[4]*Condensed Matter Theory and Computational Lab, Department of Physics, Indian Institute of Technology Madras, Chennai 600036, India*
[5]*Laboratory for Muon Spin Spectroscopy, Paul Scherrer Institute, Villigen, PSI, Switzerland*
[6]*Department of Physics, Nano Functional Materials Technology Centre and Materials Science Research Centre, Indian Institute of Technology Madras, Chennai-600036, India*
[7]*Quantum Centre of Excellence for Diamond and Emergent Materials, Indian Institute of Technology Madras, Chennai 600036, India*

†andrej.zorko@ijs.si
‡nandab@iitm.ac.in
*pkhuntia@iitm.ac.in

(Dated: 02 June, 2025)

Frustrated magnetism continues to attract significant attention due to its potential to host novel quantum many-body phenomena and associated exotic excitations that transcend existing paradigms. Herein, we present our reply to the comment on our recent thermodynamic and muon spin relaxation studies on a frustrated double perovskite, Ba$_2$MnTeO$_6$ (henceforth BMTO). Previous studies by four independent groups, including our group, suggested a trigonal space group based on single-crystal and polycrystalline samples of BMTO, while the recent comment reports a cubic space group based on polycrystalline samples. We believe that the structure is fairly intricate because of the slight variations between the two space groups, refining the crystal structure of BMTO remains an unresolved problem that needs additional high-resolution XRD and neutron diffraction studies on high-quality single crystals. It is thought, however, that structural assignments will not greatly influence any of the primary findings related to the magnetism and spin dynamics of BMTO. These consist of a magnetic phase transition at around 21 K, the observation of antiferromagnetic magnon excitations exhibiting a gap of 1.4 K beneath the phase transition, the presence of short-range spin correlations well above the antiferromagnetic phase transition, and the persistence of spin dynamics even within the magnetically ordered phase. It is important to note that the magnetization, specific heat, and μSR findings that constitute the core of our earlier study are independent; the interpretation of these findings did not rely on any specific space group. Concerning the final allocation of the symmetry of BMTO, a definitive differentiation in certain physical characteristics resulting from the symmetry is still necessary.

Three reports[1,2,3] existed on the crystal structure and magnetic properties of the perovskite material $Ba_2MnTeO_6$ (henceforth BMTO) prior to the publication of Ref. 4. The very first report[1] suggested a trigonal space group $R\bar{3}m$ based on single-crystal X-ray diffraction (XRD) results and indicated that BMTO is a triangular lattice antiferromagnet. Further work based on neutron diffraction on polycrystalline BMTO and density functional theory (DFT)[2] suggested the same trigonal space group $R\bar{3}m$ and also established that BMTO is a triangular lattice antiferromagnet. In addition, a recent report[5] on the polycrystalline samples of BMTO claimed a trigonal space group and another study[6] by an independent group on high-quality single crystals of $Ba_2MnTeO_6$ based on XRD and high-resolution TEM also suggested a trigonal space group. In contrast, Mustonen *et al.* proposed an alternative cubic crystal structure ($Fm\bar{3}m$) for BMTO, based on their neutron diffraction results on polycrystalline sample[3]. Our work (Ref. 4) focused on thermodynamics and muon spin relaxation of polycrystalline samples of BMTO, as well as DFT calculations[4], and not on determining its symmetry. The basic structural characterization of high-purity samples in the latter study suggested that BMTO crystallises in the lower-symmetry trigonal space group $R\bar{3}m$, which is consistent with the majority of previous reports[1,2] but contradicts one report[3]. It is worth noting that the sole purpose of our XRD investigation, as reported in Ref. 4, was to check the phase purity of the reported material, and the Rietveld refinement of the XRD was done following the atomic parameters published by the first paper[1] on BMTO by Wulff et al. It is not feasible to derive the exact structure based on the XRD results on polycrystalline samples. The exact and unambiguous determination of structure requires neutron diffraction experiments on high-quality single crystals, which was beyond the scope of our report[4].

Despite the assignment of different space groups for BMTO, the central theme and conclusions of the four studies[1,2,3,4] concerning the magnetism of BMTO remain intact and consistent. It is clear that Ref. 4 is neither the first nor the last paper mentioning the trigonal space group in BMTO. It is fair and relevant that fresh insights regarding the crystal structure of BMTO be provided. These should be addressed to the original paper[1] concerned with the crystal structure of BMTO, or other paper based on neutron diffraction[2] or similar microscopic techniques employed to tackle the crystal structure of BMTO.

The recent comment[7] by Mustonen *et al.* claims the cubic structure $Fm\bar{3}m$ of BMTO based on new analysis of powder neutron diffraction and suggests this has a profound effect on magnetism of this material. It is timely and important to investigate what new physics could emerge from this material by assigning different space groups (trigonal vs. cubic). The pertinent question is what this comment contributes in this context. The answer is nothing apart from idle words of how the exact symmetry is of fundamental importance in considering the properties of this material. It should be noted that the intention of our paper (Ref. 4) was not to determine the precise crystal structure or associated symmetry of BMTO by using XRD, but rather to provide the first microscopic insight into magnetism of this material via various complementary experimental techniques. The exact crystal structure, in fact, does not change the main experimental findings, i.e., observation of short-range and long-range correlations in BMTO, in any possible way.

In Ref. 4, in fact, we considered the possibility of both proposed crystal structures and concluded that the trigonal ($R\bar{3}m$) and cubic ($Fm\bar{3}m$) structures can both index the observed XRD peaks in the high-quality polycrystalline samples of BMTO, although a somewhat worse agreement with the experiment (goodness of fit $\chi^2 = 5.56$) was found for the cubic structure than for the trigonal crystal structure ($\chi^2 = 4.8$). A detailed comparison of the DFT results involving exchange couplings with a cubic model based on the data taken on high quality single crystals may shed some insights, which is beyond the scope of the present study. The exchange couplings derived from the DFT results are self-sustaining, and they are unlikely to change the main conclusion of the paper[4]. The conclusion was that the two structures are only marginally different, so that the exact structure likely has no significant effect on magnetism[2,4]. Similarly, Ref. 2 suggested that the two structures are very close in describing the material under study. According to the phase diagram as presented in Ref. 2, there is no difference in the ground state if cubic structure is considered (leading to equal nearest-neighbour intralayer and interlayer couplings) of for small trigonal distortion leading to slightly different exchanges, as reported in Ref. 4. It is also worth noting that comprehensive studies on BMTO and closely related compounds suggested that the trigonal space group ($R\bar{3}m$) offers an additional degree of freedom for the positions of Ba and O sites along the *c*-axis[2,8,9,10].

In view of the complexities involved and the tiny differences between the two structures, refining the crystal structure of BMTO is an open issue that would require further high-resolution XRD and neutron diffraction investigations on high quality single crystals. It is believed, however, that the structural refinement will in no significant way affect any of the main results regarding magnetism and spin dynamics of BMTO[4]. These include the presence of a magnetic phase transition at $T_N$ = 21 K and the presence of antiferromagnetic magnon excitations with a gap of 1.4 K below $T_N$, the presence of short-range spin correlations high above $T_N$ and the persistence of spin dynamics even in the magnetically ordered state. It is worth mentioning that the magnetization, specific heat, and μSR results that constitute the essence of Ref. 4 stand on their own, and the analysis of these results did not involve any particular space group. Regarding the ultimate assignment of the symmetry of BMTO, clear-cut distinction in some physical property arising from the symmetry is still required.

**Acknowledgements**

We thank DST, India for the PPMS facility at IIT Madras. P.K. acknowledges the funding by the Science and Engineering Research Board, and Department of Science and Technology, India through Research Grants. A.Z. acknowledges the funding by the Slovenian Research Agency through the Projects No. N1-0148, J1-2461, and Program No. P1-0125. B.R.K.N. acknowledges the funding by the Department of Science and Technology, India, through Grant No. CRG/2020/004330. BRKN acknowledges the use of the computing resources at HPCE, IIT Madras. H.L. acknowledges the financial support by SNF through the Project Grant 200021L-192109.

**Author Contributions**

All authors contributed substantially to this work.

**Competing interests**

The authors declare no competing interests.

**Correspondence** and requests for materials should be addressed to A.Z., B.R.K.N. or P.K.